\documentclass[12pt]{article}

\usepackage{amsmath,epsfig,epsf,psfrag,natbib}
\usepackage{amssymb}
\usepackage[latin1]{inputenc}

\usepackage{amsbsy}
\usepackage{lscape}
\usepackage{color,latexsym,amsfonts,amssymb,amsthm,amscd,amsmath,graphicx}
\usepackage{courier}
\usepackage{multirow}
\usepackage{booktabs}
\usepackage{hyperref}
\usepackage{ulem}

\numberwithin{equation}{section}
\numberwithin{table}{section}

\newcommand{\be}{\begin{equation} \begin{aligned}}
\newcommand{\ee}{\end{aligned} \end{equation}}
\newcommand{\bes}{\begin{equation*} \begin{aligned}}
\newcommand{\ees}{\end{aligned} \end{equation*}}
\newcommand{\bt}{\begin{tabular}}
\newcommand{\et}{\end{tabular}}
\newcommand{\ba}{\begin{eqnarray}}
\newcommand{\ea}{\end{eqnarray}}
\newcommand{\bas}{\begin{eqnarray*}}
\newcommand{\eas}{\end{eqnarray*}}
\newtheorem{theorem}{Theorem}

\newtheorem{lemma}{Lemma}
\newtheorem{proposition}{Proposition}

\newtheorem{corollary}{Corollary}

\newcommand{\balpha}{\boldsymbol\alpha}
\newcommand{\bbeta}{\boldsymbol\beta}
\newcommand{\bgamma}{\boldsymbol\gamma}

\newcommand{\bnu}{\boldsymbol \nu}

\newcommand{\bSigma}{\boldsymbol{\Sigma}}
\newcommand{\bGamma}{\boldsymbol{\Gamma}}
\newcommand{\bOmega}{\boldsymbol{\Omega}}
\newcommand{\bxi}{\boldsymbol{\xi}}
\newcommand{\bpi}{\boldsymbol{\pi}}
\newcommand{\bphi}{\boldsymbol{\varphi}}

\newcommand{\bvarphi}{\boldsymbol{\varphi}}

\newcommand{\bA}{\boldsymbol{A}}

\newcommand{\bb}{\boldsymbol{b}}

\newcommand{\bD}{\boldsymbol{D}}

\newcommand{\bO}{\mathcal{O}}

\newcommand{\bu}{\boldsymbol{u}}
\newcommand{\bV}{\boldsymbol{V}}
\newcommand{\bv}{\boldsymbol{v}}

\newcommand{\bM}{\boldsymbol{M}}

\newcommand{\bq}{\boldsymbol q}

\newcommand{\bI}{\mbox{\bf I}}

\newcommand{\0}{\mbox{\bf 0}}
\newcommand{\1}{\mbox{\bf 1}}

\newcommand{\E}{\mathbb{E}}

\allowdisplaybreaks[4]

\begin{document}

\title{Semiparametric inference for inequality measures under nonignorable nonresponse using callback data
}
\date{}
\maketitle

\maketitle
\vspace{-1in}
\begin{center}

{Xinyu Wang\\
 Department of Statistics and Data Science, School of Economics, Xiamen University, China
}
{Chunlin Wang\\
 Department of Statistics and Data Science, School of Economics and Wang Yanan Institute for Studies in Economics, Xiamen University, China\\
}
{Tao Yu \\
Department of Statistics and Data Science, National University of Singapore, Singapore
}
{Pengfei Li \\
Department of Statistics and Actuarial Science,
University of Waterloo, Canada
}

\end{center}

\begin{abstract}
\noindent
This paper develops semiparametric methods for estimation and inference of widely used inequality measures when survey data are subject to nonignorable nonresponse, a challenging setting in which response probabilities depend on the unobserved outcomes. Such nonresponse mechanisms are common in household surveys and invalidate standard inference procedures due to selection bias and lack of population representativeness. We address this problem by exploiting callback data from repeated contact attempts and adopting a semiparametric model that leaves the outcome distribution unspecified.
We construct semiparametric full-likelihood estimators for the underlying distribution and the associated inequality measures, and establish their large-sample properties for a broad class of functionals, including quantiles, the Theil index, and the Gini index. Explicit asymptotic variance expressions are derived, enabling valid Wald-type inference under nonignorable nonresponse. To facilitate implementation, we propose a stable and computationally convenient expectation-maximization algorithm, whose steps either admit closed-form expressions or reduce to fitting a standard logistic regression model. Simulation studies demonstrate that the proposed procedures effectively correct nonresponse bias and achieve near-benchmark efficiency. An application to Consumer Expenditure Survey data illustrates the practical gains from incorporating callback information when making inference on inequality measures.\\

\end{abstract}


\section{Introduction}
\label{sec1}

In economics and official statistics, inequality measures play a central role in summarizing the distributions of economic variables such as income, wealth, and expenditure \citep{Atkinson1970}.
Let $F(y)$ denote the cumulative distribution function (CDF) of a continuous random variable $Y$ supported on $(0,\infty)$, which may represent, for example, household income.
In this paper, we focus on commonly used inequality measures expressed as  $\theta(F)$, where $\theta(\cdot)$ is  a statistical functional of $F(y)$. These measures include the Gini index, quantiles, the Theil index, the coefficient of variation, and members of the generalized entropy and Atkinson classes (see Section \ref{Sec2.2} for their formal definitions). For a comprehensive review of inequality measures and their theoretical properties, see \cite{Cowell2011}.

There is a vast literature on statistical inference for inequality measures; see \cite{Davidson2007,Giorgi2017,Dufour2019,Yuan2023,Zalghout2025} and the references therein.
Much of this literature, however, implicitly assumes that the observed data are representative of the target population. This assumption is often violated in practice, as survey samples used to measure inequality frequently suffer from nonresponse or missing data; see \cite{Groves1998}.

According to \citet{Rubin1976}, valid inference for $\theta(F)$ must account for the underlying missing-data mechanism.
If the response probability depends on the unobserved value of $Y$, the mechanism is said to be \emph{nonignorable}; otherwise, it is ignorable.
Nonignorable nonresponse mechanisms are common in surveys on income and other sensitive information \citep{Korinek2006,Korinek2007,Bollinger2013,Bollinger2019}, as high-income households are often less likely to respond.
Under such mechanisms, the observed data constitute a biased sample of the target population, because the income distribution of respondents may differ systematically from the true distribution $F(y)$ \citep{Qin2017}.
Moreover, $F(y)$ is not nonparametrically identifiable based solely on the biased sample \citep{Wang2014,Miao2016,Morikawa2021}.
Consequently, many standard inference procedures for $\theta(F)$ are either inapplicable or yield inefficient and potentially misleading results \citep{Paraje2010,Alvarez2021}.
In general, without additional assumptions or auxiliary information, valid statistical inference for $\theta(F)$ is not possible when only a sample subject to nonignorable nonresponse is available.

In survey practice, additional contact attempts are commonly made when the initial attempt fails.
Such callback strategies are widely used to increase response rates and to monitor survey quality \citep{DrewFuller1980,Couper1998,Bates2000,Durrant2012,Kreuter2013,Olson2013}.
The resulting callback data, which record the number of contact attempts for initially nonresponding units, are routinely collected and provide valuable auxiliary information for identifying $F(y)$ and for modeling and adjusting for nonignorable nonresponse \citep{Alho1990,Qin2014,Miao2024}.

Let $m \ge 2$ denote the maximum number of contact attempts, and let $D$ be the callback variable, where $D=j$ if a household responds at the $j$th call, $j=1,\ldots,m$, and $D=m+1$ if it never responds. Thus, $Y$ is observed when $D \le m$ and missing otherwise.
Define $\pi_j(y)=\Pr(D=j \mid Y=y, D \ge j)$ as the response probability at the $j$th call given nonresponse to earlier calls. The functions $\pi_j(y)$ characterize the nonignorable missing-data mechanism.
Because income distributions are typically skewed and may exhibit complex shapes, we adopt the semiparametric model of \citet{Alho1990}, which leaves $F(y)$ unspecified while modeling each $\pi_j(y)$ parametrically via a logistic regression,
\begin{equation}
\label{Alho}
\pi_j(y)
= \frac{\exp\{\alpha_j+\bbeta^\top \bq(y)\}}
{1+\exp\{\alpha_j+\bbeta^\top \bq(y)\}},
\quad j=1,\ldots,m.
\end{equation}
Here, $\bq(\cdot)$ is a prespecified $d$-dimensional function, with parameters $\balpha=(\alpha_1,\ldots,\alpha_m)^\top$ and $\bbeta$.
When $\bq(y)=y$, this reduces to Alho's original model, while allowing a flexible choice of $\bq(y)$ improves adaptability.
In this formulation, $\alpha_j$ captures call-specific response heterogeneity, and $\bbeta$ is common across calls under the continuum of resistance assumption, which ensures identifiability of the semiparametric model \citep{Guan2018,Miao2024}.

Under model \eqref{Alho}, \citet{Alho1990} proposed a conditional likelihood approach to estimate $\balpha$ and $\bbeta$, leading to an inverse probability weighting (IPW) estimator of the population mean.
Building on the same model, \citet{Qin2014} developed a semiparametric full-likelihood method that estimates $\balpha$, $\bbeta$, and the population mean.
Their approach yields efficient inference that is robust to assumptions on $F(y)$ and avoids the instability of IPW estimators \citep{Li2023}, particularly when the estimated response probabilities are small.
Recent extensions using callback data include generalized method of moments estimation of $\balpha$, $\bbeta$, and the finite population mean \citep{Kim2014}, sensitivity analysis \citep{Daniels2015}, and the incorporation of covariate information \citep{Guan2018,Chen2018}.

To the best of our knowledge, no existing methods use callback data to develop valid inference procedures for inequality measures $\theta(F)$ under nonignorable nonresponse.
Existing results based on callback data focus primarily on inference for the mean of $Y$ and do not extend to more general measures $\theta(F)$, which may involve complex linear, nonlinear, or nonsmooth functionals of $F$, such as quantiles or the Gini index.
Unlike inference for the mean, inference for these measures is substantially more challenging, because establishing the theoretical properties of the associated estimators requires delicate asymptotic arguments.
Moreover, existing methods rely on complex numerical optimization, and no efficient algorithm with theoretical guarantees is currently available.

In this paper, we develop efficient and reliable inference procedures for general inequality measures $\theta(F)$ under nonignorable nonresponse by incorporating callback data within the semiparametric model of \citet{Alho1990}.
Our contributions are summarized as follows.

\begin{itemize}
\item[(a)]
We propose semiparametric maximum full-likelihood estimators for inequality measures $\theta(F)$ that adjust for sample selection bias arising from nonignorable nonresponse.

\item[(b)]
We establish the asymptotic properties of the proposed estimators and derive explicit analytical expressions for their asymptotic variances.
In particular, we use the Bahadur representation to analyze estimators of quantiles and their functions, and apply the theory of U- and V-statistics to study the Gini index estimator.
These results lead to valid Wald-type confidence intervals for a broad class of inequality measures.

\item[(c)]
We develop a novel and stable expectation-maximization (EM) algorithm for numerical implementation.
The conditional expectations in the E-step admit closed-form expressions, and the constrained maximization in the M-step reduces to fitting a logistic regression model using standard software.
We also establish the monotonicity property of the algorithm.

\item[(d)]
Through simulation studies and a real-data application, we demonstrate the favorable finite-sample performance of the proposed methods.
\end{itemize}

The paper is organized as follows.
Section~\ref{Sec_estimation} proposes full-likelihood estimators for the model parameters and $\theta(F)$, establishes their asymptotic properties, and constructs valid confidence intervals for various inequality measures.
Section~\ref{sec.EM} presents an EM algorithm for practical implementation.
Section~\ref{Sec_simulation} reports the results of simulation studies, and Section~\ref{Sec_real} analyzes a real income data set from the Consumer Expenditure Survey.
Section~\ref{Sec_discussion} concludes with some remarks.
All proofs are provided in the Supplementary Material.

\section{Semiparametric estimation and inference of inequality measures} 
\label{Sec_estimation}

\subsection{Semiparametric full-likelihood approach} 
\label{Sec2.1}

Suppose we observe a random sample $\{(Y_i, D_i): i=1,\ldots,N\}$ generated from Alho's semiparametric model~\eqref{Alho}, where $F(y)$ is left unspecified.
Let $n = \sum_{i=1}^N I(D_i \le m)$ denote the random number of respondents.
Without loss of generality, we index the first $n$ observations as respondents and the remaining $N-n$ as nonrespondents, for whom the $Y_i$'s are unobserved.
Hereafter, we refer to the first $n$ observations as the complete-case (CC) data.

Let
\bes
\rho_1(y;\balpha,\bbeta)
&= \Pr(D=1 \mid Y=y) = \pi(y;\alpha_1,\bbeta), \\
\rho_j(y;\balpha,\bbeta)
&= \Pr(D=j \mid Y=y)
= \pi(y;\alpha_j,\bbeta)\prod_{i=1}^{j-1}\{1-\pi(y;\alpha_i,\bbeta)\},
\quad j=2,\ldots,m, \\
\rho(y;\balpha,\bbeta)
&= \Pr(D \le m \mid Y=y)
= \sum_{j=1}^m \rho_j(y;\balpha,\bbeta).
\ees
Then
\[
\eta = \Pr(D \le m) = \int_0^\infty \rho(y;\balpha,\bbeta)\, dF(y)
\]
denotes the marginal probability that a generic household responds within $m$ callbacks.
Denote the full parameter vector as $\bvarphi = (\balpha,\bbeta,\eta,F)$.

We now derive the full likelihood for $\bvarphi$.
It factors into three components:
(i) the likelihood of $D_1,\ldots,D_n$ conditional on $Y_1,\ldots,Y_n$ and on the event that the $n$ households respond within $m$ callbacks;
(ii) the likelihood of $Y_1,\ldots,Y_n$ given that these $n$ households respond within $m$ callbacks; and
(iii) the likelihood of the total number of respondents $n$.

First, given that the $i$th household responded within $m$ callbacks and provided the observed response $Y_i$, the conditional probability of responding on the $D_i$th callback is
\begin{align*}
\Pr(D = D_i \mid Y = Y_i, D_i \le m)
&= \frac{\Pr(D = D_i, D_i \le m \mid Y = Y_i)}{\Pr(D_i \le m \mid Y = Y_i)} \\
&= \frac{\Pr(D = D_i \mid Y = Y_i)}{\Pr(D_i \le m \mid Y = Y_i)} \\
&= \frac{\prod_{j=1}^m \{\rho_j(Y_i;\balpha,\bbeta)\}^{I(D_i=j)}}{\rho(Y_i;\balpha,\bbeta)}.
\end{align*}
Hence, the likelihood contribution from $D_1,\ldots,D_n$ conditional on $Y_1,\ldots,Y_n$ and on the event that the $n$ households responded within $m$ callbacks, also known as the conditional likelihood \citep{Alho1990}, is
\begin{equation}
\label{like.part1}
L_1
= \prod_{i=1}^n
\frac{\prod_{j=1}^m \{\rho_j(Y_i;\balpha,\bbeta)\}^{I(D_i=j)}}
{\rho(Y_i;\balpha,\bbeta)}.
\end{equation}

Second, given that the $i$th household responded within $m$ callbacks, the conditional probability of observing $Y_i$ is
\begin{equation}
\label{part2.post}
\Pr(Y = Y_i \mid D_i \le m)
= \frac{\Pr(D_i \le m \mid Y = Y_i)\Pr(Y = Y_i)}{\Pr(D_i \le m)}
= \frac{\rho(Y_i;\balpha,\bbeta)\, dF(Y_i)}{\eta}.
\end{equation}
Therefore, the likelihood contribution from $Y_1,\ldots,Y_n$ given that the $n$ households responded within $m$ callbacks is
\begin{equation}
\label{like.part2}
L_2
= \prod_{i=1}^n \frac{\rho(Y_i;\balpha,\bbeta)\, dF(Y_i)}{\eta}.
\end{equation}

Third, the number of respondents $n$ follows a $\mathrm{Binomial}(N,\eta)$ distribution, with likelihood
\begin{equation}
\label{like.part3}
L_3
= {N \choose n}\eta^n(1-\eta)^{N-n}.
\end{equation}

Combining \eqref{like.part1} and \eqref{like.part2}--\eqref{like.part3} yields the full likelihood of $\bvarphi$ as
\begin{equation*}
L = L_1 L_2 L_3
= {N \choose n}
\prod_{i=1}^n \prod_{j=1}^m \{\rho_j(Y_i;\balpha,\bbeta)\}^{I(D_i=j)}
\cdot \prod_{i=1}^n dF(Y_i)
\cdot (1-\eta)^{N-n}.
\end{equation*}
Following the empirical likelihood principle \citep{Owen2001}, we model $F$ as
\[
F(y) = \sum_{i=1}^{n} p_i I(Y_i \le y),
\]
where $p_i = dF(Y_i)$ for $i=1,\ldots,n$.
Up to an additive constant that does not depend on $\bvarphi$, the semiparametric full log-likelihood function is
\be
\label{log_lik}
\ell(\bvarphi)
= \sum_{i=1}^{n} \sum_{j=1}^m I(D_i=j)\log\{\rho_j(Y_i;\balpha,\bbeta)\}
+ \sum_{i=1}^{n}\log(p_i)
+ (N-n)\log(1-\eta),
\ee
where $\bvarphi$ is subject to the constraints
\be
\label{const}
\mathcal{C}
= \left\{
\bvarphi :
p_i > 0,\;
\sum_{i=1}^{n} p_i = 1,\;
\sum_{i=1}^{n} p_i\{\rho(Y_i;\balpha,\bbeta)-\eta\} = 0
\right\}.
\ee
The first two constraints ensure that $F$ is a valid CDF, while the last follows from the definition of $\eta$ and accounts for sample selection bias induced by nonignorable nonresponse.

The semiparametric maximum full-likelihood estimator of $\bvarphi$ is defined as
\be
\label{hatphi}
\hat{\bvarphi}
= \arg\max_{\bvarphi \in \mathcal{C}} \ell(\bvarphi).
\ee
Given $\hat{\bvarphi}$, the corresponding semiparametric maximum full-likelihood estimator of $F(y)$ is
\be
\label{hatF}
\hat{F}(y)
= \sum_{i=1}^{n} \hat{p}_i I(Y_i \le y).
\ee

Since a closed-form expression for $\hat{\bvarphi}$ is generally unavailable, we rely on numerical methods.
In Section~\ref{sec.EM}, we develop a stable and computationally convenient EM algorithm to compute $\hat{\bvarphi}$ and $\hat{F}(y)$.
The resulting estimator $\hat{F}(y)$ provides the foundation for correcting sample selection bias and for estimating the inequality measures $\theta(F)$ in the presence of nonignorable nonresponse.

\subsection{Semiparametric estimation of inequality measures}
\label{Sec2.2}

In this section, we propose a semiparametric approach for estimating inequality measures.
Recall that the measures of interest, $\theta(F)$, can be expressed as statistical functionals of the income distribution $F$.
Given the estimator $\hat{F}(y)$, a natural strategy is to adopt the plug-in principle by replacing $F(y)$ with $\hat{F}(y)$, which yields $\theta(\hat{F})$ as the semiparametric estimator of $\theta(F)$. This differs fundamentally from the IPW-type approach.
For simplicity, we denote the resulting semiparametric full-likelihood estimator by $\hat{\theta} = \theta(\hat{F})$.
We focus on three widely used classes of inequality measures, each corresponding to a specific functional form of $\theta(\cdot)$.

\subsubsection{Quantiles and their functions} 

In applications, population quantiles and their functions are widely recognized as important measures of inequality \citep{Prendergast2017,Zalghout2025}.
Common examples include the median, quartiles, quintiles, percentiles, and the interquartile range of income.
Moreover, following the implementation of new policies, trends in quantiles, as well as their differences or ratios, provide valuable information for monitoring changes in the income distribution and assessing policy impacts.
In this section, we focus on the estimation of quantiles and their functions, which are nonlinear and nonsmooth functionals of $F$.

For $\tau \in (0,1)$, the $100\tau$th \emph{quantile} of $F(y)$ is defined as
\bes
\theta_{\mathrm{quan}}^{\tau}(F)
= F^{-1}(\tau)
:= \inf\{y : F(y) \ge \tau\}.
\ees
Replacing $F(y)$ with $\hat{F}(y)$ in \eqref{hatF}, the proposed semiparametric estimator of $\theta_{\mathrm{quan}}^{\tau}(F)$ is given by
\bes
\hat{\theta}_{\mathrm{quan}}^{\tau}
= \theta_{\mathrm{quan}}^{\tau}(\hat{F})
= \inf\{y : \hat{F}(y) \ge \tau\},
\quad \tau \in (0,1).
\ees
If interest lies in functions of quantiles, such as the quantile difference
$\theta_{\mathrm{quan}}^{\tau_1} - \theta_{\mathrm{quan}}^{\tau_2}$
or the quantile ratio
$\theta_{\mathrm{quan}}^{\tau_1}/\theta_{\mathrm{quan}}^{\tau_2}$
for $\tau_1,\tau_2 \in (0,1)$, the corresponding semiparametric estimators are
$\hat{\theta}_{\mathrm{quan}}^{\tau_1} - \hat{\theta}_{\mathrm{quan}}^{\tau_2}$
and
$\hat{\theta}_{\mathrm{quan}}^{\tau_1}/\hat{\theta}_{\mathrm{quan}}^{\tau_2}$,
respectively.

\subsubsection{A class of linear functionals and its functions}

We next consider an important class of inequality measures that can be expressed as linear functionals of $F$ and their transformations.
A $p$-dimensional linear functional of $F$ is defined as
\be
\label{gamma}
\bgamma_{\bu}
= \int_0^\infty \bu(y)\, dF(y),
\ee
for suitable choices of $\bu(y) = \big(u_1(y),\ldots,u_p(y)\big)^\top$.
Many commonly used inequality measures can be written in the unified form
\be
\label{hgamma}
\theta_{\bu,h}(F)
= h(\bgamma_{\bu}),
\ee
where $h(\cdot): \mathbb{R}^p \to \mathbb{R}$ is a smooth function of $\bgamma_{\bu}$.
Examples in this class include population moments (centered or uncentered), the Theil index, the coefficient of variation, the mean logarithmic deviation, and members of the generalized entropy and Atkinson classes.
Table~\ref{tabtab} summarizes these measures along with the corresponding specifications of $\bu(\cdot)$ and $h(\cdot)$.

\begin{table}[!ht]
\tabcolsep 6pt
\renewcommand{\arraystretch}{1}
\caption{Examples of inequality measures in the class $\theta_{\bu,h}(F)$ with specific forms of $\bu(\cdot)$ in \eqref{gamma} and $h(\cdot)$ in \eqref{hgamma}.}
\label{tabtab}
\centering
\begin{tabular}{l cc}
\toprule
Inequality measure & $\bu(y)$ & $h(\cdot)$ \\
\midrule
$k$th moment & $y^k$ & $h(x)=x$ \\
$k$th centered moment
& $(1,y,\ldots,y^k)^\top$
& $h(x_0,x_1,\ldots,x_k)=\sum_{s=0}^k {k\choose s} x_s (-x_1)^{k-s}$ \\
coefficient of variation
& $(y, y^2)^\top$
& $h(x_1,x_2)=\sqrt{x_2 - x_1^2}/x_1$ \\
generalized entropy class
& $(y, y^{\zeta})^\top$
& $\displaystyle h(x_1,x_2)=\frac{1}{\zeta(\zeta-1)}\left(x_2/x_1^{\zeta}-1\right),\ \zeta\in(0,1)$ \\
Theil index
& $(y, y\log(y))^\top$
& $h(x_1,x_2)=x_2/x_1-\log(x_1)$ \\
mean logarithmic deviation
& $(y, \log(y))^\top$
& $h(x_1,x_2)=\log(x_1)-x_2$ \\
Atkinson class
& $(y, y^\zeta)^\top$
& $h(x_1,x_2)=1-x_2^{1/\zeta}/x_1,\ \zeta\in(0,1)$ \\
\bottomrule
\end{tabular}
\end{table}

In our numerical studies, we focus on an important and widely used inequality measure, the \emph{Theil index} \citep{Theil1967}, defined as
\bes
\theta_{\mathrm{Theil}}(F)
= \frac{\int_0^\infty y\log(y)\, dF(y)}{\int_0^\infty y\, dF(y)}
- \log\!\left\{\int_0^\infty y\, dF(y)\right\}.
\ees
It is immediate that $\theta_{\mathrm{Theil}}(F)$ is a special case of the class $\theta_{\bu,h}(F)$ with
$\bu(y) = \big(y, y\log(y)\big)^\top$ and $h(x_1,x_2)=x_2/x_1-\log(x_1)$.
Moreover, $\theta_{\mathrm{Theil}}(F)$ corresponds to the limiting case of the generalized entropy class as the governing parameter $\zeta \to 1$.
A useful property of the Theil index, and more generally of functionals of the form \eqref{hgamma}, is their additive decomposability by subgroup \citep{Cowell2011}, which facilitates the analysis of within-group and between-group inequality.

By definition \eqref{hgamma}, $\theta_{\bu,h}(F)$ can be estimated by replacing $F(y)$ with $\hat{F}(y)$ in \eqref{hatF}.
The proposed semiparametric estimator for this class of inequality measures is
\[
\hat{\theta}_{\bu,h}
= \theta_{\bu,h}(\hat{F})
= h(\hat{\bgamma}_{\bu}),
\quad
\text{with}
\quad
\hat{\bgamma}_{\bu}
= \sum_{i=1}^n \hat{p}_i \bu(Y_i).
\]
In particular, the proposed estimator of the Theil index is
\bes
\hat{\theta}_{\mathrm{Theil}}
= \frac{\sum_{i=1}^n \hat{p}_i Y_i \log(Y_i)}{\sum_{i=1}^n \hat{p}_i Y_i}
- \log\!\left(\sum_{i=1}^n \hat{p}_i Y_i\right).
\ees

\subsubsection{Gini index}

Although $\theta_{\mathrm{quan}}^{\tau}(F)$ and $\theta_{\bu,h}(F)$ cover a wide range of inequality measures, the Gini index \citep{Gini1912} is a notable exception that cannot be expressed within these classes.
In this section, we therefore consider the Gini index through a specific functional form of $\theta(\cdot)$.
The \emph{Gini index} has several equivalent representations; here, we adopt the following formulation \citep{Yuan2023}:
\begin{equation}
\label{Gini}
\theta_{\mathrm{Gini}}(F)
= \frac{\psi(F)}{\mu(F)} - 1,
\end{equation}
where $\psi(F) = \int_0^\infty 2y F(y)\, dF(y)$ and $\mu(F) = \int_0^\infty y\, dF(y)$.
Unlike the measures considered previously, $\theta_{\mathrm{Gini}}(F)$ is distinct in that $\psi(F)$ involves a U-statistic-type functional of $F$.
The Gini index is among the most widely used measures of inequality and is routinely reported by major official statistical agencies worldwide.
Its widespread use is also attributable to its close relationship with the Lorenz curve \citep{Lorenz1905}, which provides an intuitive graphical interpretation.

Using $\hat{F}(y)$ in \eqref{hatF}, the proposed semiparametric estimator of $\theta_{\mathrm{Gini}}(F)$ is
$\hat{\theta}_{\mathrm{Gini}} = \theta_{\mathrm{Gini}}(\hat{F})$, which can be written explicitly as
\be
\label{Gini.est}
\hat{\theta}_{\mathrm{Gini}}
= \frac{2\sum_{i=1}^{n} \hat{p}_i Y_i \hat{F}(Y_i)}{\sum_{i=1}^{n} \hat{p}_i Y_i} - 1
= \frac{\hat{\psi}}{\hat{\mu}} - 1,
\ee
where $\hat{\mu} = \mu(\hat{F}) = \sum_{i=1}^{n} \hat{p}_i Y_i$ and
$\hat{\psi} = \psi(\hat{F}) = 2\sum_{i=1}^{n} \hat{p}_i Y_i \hat{F}(Y_i)$.
These quantities are introduced for notational convenience in subsequent theoretical developments.

\subsection{Asymptotic properties}
\label{Sec2.3}

In this section, we investigate the asymptotic behavior of the estimators of inequality measures proposed in Section~\ref{Sec2.2}.
Although the plug-in principle provides a natural approach for estimating inequality measures $\theta(F)$, establishing the asymptotic properties of the resulting estimators for statistical inference is nontrivial and requires a careful combination of appropriate asymptotic tools.

To facilitate the presentation, we introduce additional notation.
Let $(\balpha^*, \bbeta^*, \eta^*)$ denote the true value of the model parameters $(\balpha, \bbeta, \eta)$.
Define $\rho^*(y) = \rho(y; \balpha^*, \bbeta^*)$ and
$\bpi^*(y) = \big(\pi(y;\alpha_1^*, \bbeta^*), \ldots, \pi(y;\alpha_m^*, \bbeta^*)\big)^\top$.
Let $\1_k$ and $\0_k$ denote $k \times 1$ vectors of ones and zeros, respectively, and let $\bI_k$ denote the $k \times k$ identity matrix.
We further define
\[
\bA(y) = \big(\bI_m,\ \1_m \bq(y)^\top\big)^\top
\quad \text{and} \quad
\bv(y)
= \rho^*(y)^{-1}
\big(\{\rho^*(y) - 1\}\bpi^*(y)^\top \bA(y)^\top,\ 1,\ (\eta^*)^2\big)^\top.
\]

We begin by studying the asymptotic behavior of the proposed semiparametric estimator $\hat{F}(y)$ of $F(y)$.

\begin{theorem}
\label{Thm_F}
Suppose that Conditions C1--C5 in the Appendix are satisfied.
Under model~\eqref{Alho}, as $N \to \infty$,
\(
N^{1/2}\big\{\hat{F}(y) - F(y)\big\}
\)
converges weakly to a zero-mean tight Gaussian process with continuous sample paths and covariance function
\bes
\Sigma(y_1, y_2)
&= \E\!\left\{\frac{\xi_F(Y, y_1)\xi_F(Y, y_2)}{\rho^*(Y)}\right\}
+ \E\{\xi_F(Y, y_1)\bv(Y)\}^{\top}
\bGamma
\E\{\xi_F(Y, y_2)\bv(Y)\},
\ees
for any $y_1, y_2 \in (0,\infty)$, where
$\xi_F(Y,y) = I(Y \le y) - F(y)$ and $\bGamma$ is defined in \eqref{Gamma} in the Appendix.
\end{theorem}

The result of Theorem~\ref{Thm_F} provides the foundation for deriving the asymptotic properties of the proposed estimators of inequality measures $\hat{\theta}_{\mathrm{quan}}^{\tau}$, $\hat{\theta}_{\bu,h}$, and $\hat{\theta}_{\mathrm{Gini}}$ in the subsequent theorems.

\subsubsection{Asymptotic properties of $\hat{\theta}_\text{quan}^{\tau}$} 

Building on Theorem~\ref{Thm_F}, we first apply the von Mises calculus and the functional delta method to derive a Bahadur representation for $\hat{\theta}_{\mathrm{quan}}^{\tau}$ \citep{Bahadur1966}.
This representation provides a convenient and tractable route for establishing the asymptotic properties of $\hat{\theta}_{\mathrm{quan}}^{\tau}$.
Let $\theta_{\mathrm{quan}}^{\tau *}$ denote the true value of $\theta_{\mathrm{quan}}^{\tau}(F)$.

\begin{lemma}
\label{Bahadur}
Assume that the conditions of Theorem~\ref{Thm_F} are satisfied.
For a given $\tau \in (0,1)$, the estimator $\hat{\theta}_{\mathrm{quan}}^{\tau}$ admits the following Bahadur representation:
\bes
\hat{\theta}_{\mathrm{quan}}^{\tau} - \theta_{\mathrm{quan}}^{\tau *}
= \frac{F(\theta_{\mathrm{quan}}^{\tau *}) - \hat{F}(\theta_{\mathrm{quan}}^{\tau *})}
{f(\theta_{\mathrm{quan}}^{\tau *})}
+ o_p(N^{-1/2}),
\ees
where $f(y)$ denotes the probability density function of $Y$, which is assumed to be positive and continuously differentiable in a neighborhood of $\theta_{\mathrm{quan}}^{\tau *}$.

\end{lemma}

Combining the Bahadur representation of $\hat{\theta}_{\mathrm{quan}}^{\tau}$ in Lemma~\ref{Bahadur} with the result of Theorem~\ref{Thm_F}, we establish the joint asymptotic normality of $\hat{\theta}_{\mathrm{quan}}^{\tau_1}$ and $\hat{\theta}_{\mathrm{quan}}^{\tau_2}$ for any $\tau_1, \tau_2 \in (0,1)$ in the following theorem.
Throughout, we use ``$\overset{d}{\to}$'' to denote convergence in distribution.

\begin{theorem}
\label{Thm_xi}
Assume that the conditions of Lemma~\ref{Bahadur} hold at levels
$\tau_1, \tau_2 \in (0,1)$, with the corresponding true quantiles
$\theta_{\mathrm{quan}}^{\tau_1 *}$ and $\theta_{\mathrm{quan}}^{\tau_2 *}$.
Under model~\eqref{Alho}, as $N \to \infty$,
\[
N^{1/2}
\big(
\hat{\theta}_{\mathrm{quan}}^{\tau_1} - \theta_{\mathrm{quan}}^{\tau_1 *},
\hat{\theta}_{\mathrm{quan}}^{\tau_2} - \theta_{\mathrm{quan}}^{\tau_2 *}
\big)^\top
\overset{d}{\to}
\mathcal{N}\!\left(\0,\; \bOmega(\tau_1,\tau_2)\right),
\]
where
\[
\bOmega(\tau_1,\tau_2)
=
\bD^{-1}
\bSigma\!\left(\theta_{\mathrm{quan}}^{\tau_1 *},
\theta_{\mathrm{quan}}^{\tau_2 *}\right)
\bD^{-1},
\]
with
\[
\bD
=
\begin{pmatrix}
f(\theta_{\mathrm{quan}}^{\tau_1 *}) & 0 \\
0 & f(\theta_{\mathrm{quan}}^{\tau_2 *})
\end{pmatrix}
\quad \text{and} \quad
\bSigma(y_1,y_2)
=
\begin{pmatrix}
\Sigma(y_1,y_1) & \Sigma(y_1,y_2) \\
\Sigma(y_1,y_2) & \Sigma(y_2,y_2)
\end{pmatrix}.
\]

\end{theorem}

Theorem~\ref{Thm_xi} also implies the marginal asymptotic normality of
$\hat{\theta}_{\mathrm{quan}}^{\tau}$.
In studies of income inequality, researchers are often interested in smooth functions of quantiles at different levels, such as
$\hat{\theta}_{\mathrm{quan}}^{\tau_1} - \hat{\theta}_{\mathrm{quan}}^{\tau_2}$
and
$\hat{\theta}_{\mathrm{quan}}^{\tau_1} / \hat{\theta}_{\mathrm{quan}}^{\tau_2}$.
Using the result of Theorem~\ref{Thm_xi}, the asymptotic distributions of such smooth functions can be conveniently derived via the delta method, as stated in the following corollary.

\begin{corollary}
\label{Q_function}
Assume that the conditions of Theorem~\ref{Thm_xi} hold for
$\theta_{\mathrm{quan}}^{\tau_1 *}$ and $\theta_{\mathrm{quan}}^{\tau_2 *}$.
Let $g(\cdot,\cdot)$ be a twice continuously differentiable function.
Then, as $N \to \infty$,
\[
N^{1/2}
\big\{
g(\hat{\theta}_{\mathrm{quan}}^{\tau_1}, \hat{\theta}_{\mathrm{quan}}^{\tau_2})
-
g(\theta_{\mathrm{quan}}^{\tau_1 *}, \theta_{\mathrm{quan}}^{\tau_2 *})
\big\}
\overset{d}{\to}
\mathcal{N}\!\left(0,\; \sigma_{\mathrm{quan}}^2(\tau_1,\tau_2)\right),
\]
where
\[
\sigma_{\mathrm{quan}}^2(\tau_1,\tau_2)
=
\bb_g^\top \bOmega(\tau_1,\tau_2) \bb_g,
\quad \text{with} \quad
\bb_g
=
\left(
\frac{\partial g (\theta_{\mathrm{quan}}^{\tau_1 *},\, \theta_{\mathrm{quan}}^{\tau_2 *})}{\partial \theta_{\mathrm{quan}}^{\tau_1}},
\frac{\partial g (\theta_{\mathrm{quan}}^{\tau_1 *},\, \theta_{\mathrm{quan}}^{\tau_2 *})}{\partial \theta_{\mathrm{quan}}^{\tau_2}}
\right)^\top.
\]
\end{corollary}

\subsubsection{Asymptotic properties of $\hat\theta_{\bu,h}$} 
We next study the asymptotic properties of the proposed estimators of inequality measures in the class
$\hat{\theta}_{\bu,h} = h(\hat{\bgamma}_{\bu})$ for smooth functions $h$, which includes the Theil index as a special case.
Let $\bgamma_{\bu}^*$ and $\theta_{\bu,h}^*$ denote the true values of $\bgamma_{\bu}$ and $\theta_{\bu,h}(F)$ defined in
\eqref{gamma} and \eqref{hgamma}, respectively.

\begin{theorem}
\label{Thm_ineq}
Suppose that Conditions C1--C5 in the Appendix are satisfied and that
$\int \rho^*(y)^{-1}\bu(y)\bu(y)^\top \, dF(y)$ is finite.
Under model~\eqref{Alho}, as $N \to \infty$,
\bes
N^{1/2}\big(\hat{\theta}_{\bu,h} - \theta_{\bu,h}^*\big)
\overset{d}{\to}
\mathcal{N}\big(0, \sigma_{\bu,h}^2\big),
\ees
where $\sigma_{\bu,h}^2 = \bb_h^\top \bSigma_{\bu} \bb_h$, with
$\bb_h = \partial h(\bgamma_{\bu}^*) / \partial \bgamma_{\bu}$,
\[
\bSigma_{\bu}
=
\E\!\left\{\frac{\bxi_{\bu}(Y)\bxi_{\bu}^\top(Y)}{\rho^*(Y)}\right\}
+
\E\!\big\{\bxi_{\bu}(Y)\bv^\top(Y)\big\}
\bGamma
\E\!\big\{\bv(Y)\bxi_{\bu}^\top(Y)\big\},
\]
and $\bxi_{\bu}(Y) = \bu(Y) - \bgamma_{\bu}^*$.
\end{theorem}

Theorem~\ref{Thm_ineq} immediately implies the asymptotic normality of the proposed estimator
$\hat{\theta}_{\mathrm{Theil}}$.
Let $\theta_{\mathrm{Theil}}^*$ denote the true value of the Theil index $\theta_{\mathrm{Theil}}(F)$.

\begin{corollary}
\label{Thm_Theil}
Assume that the conditions of Theorem~\ref{Thm_ineq} hold for
$\bu(y) = \big(y, y\log(y)\big)^\top$.
Then, as $N \to \infty$,
\[
N^{1/2}\big(\hat{\theta}_{\mathrm{Theil}} - \theta_{\mathrm{Theil}}^*\big)
\overset{d}{\to}
\mathcal{N}\big(0, \sigma_{\mathrm{Theil}}^2\big),
\]
where $\sigma_{\mathrm{Theil}}^2 = \bb_T^\top \bSigma_{\bu} \bb_T$ with
\[
\bb_T
=
\frac{1}{\gamma_1^*}
\big(-\gamma_2^*/\gamma_1^* - 1,\; 1\big)^\top,
\quad
\gamma_1^* = \int_0^\infty y\, dF(y), 
\quad
\gamma_2^* = \int_0^\infty y\log(y)\, dF(y).
\]
\end{corollary}

\subsubsection{Asymptotic properties of $\hat{\theta}_\text{Gini}$}

We now investigate the asymptotic properties of the proposed estimator of the Gini index,
$\hat{\theta}_{\mathrm{Gini}}$, which requires additional asymptotic tools beyond those used in the previous section.
Recall the definition of $\hat{\theta}_{\mathrm{Gini}}$ in \eqref{Gini.est}, where the numerator can be written as
\[
\hat{\psi}
= 2 \sum_{i=1}^{n} \hat{p}_i Y_i \hat{F}(Y_i)
= \sum_{i=1}^{n} \sum_{j=1}^{n} 2 \hat{p}_i \hat{p}_j Y_i I(Y_j \le Y_i).
\]
Unlike $\hat{\bgamma}_{\bu}$, the statistic $\hat{\psi}$ does not admit a linear representation, as it involves pairs of observations.
As a result, Theorem~\ref{Thm_ineq} is not directly applicable for deriving the asymptotic distribution of
$\hat{\theta}_{\mathrm{Gini}}$.
Instead, noting that $\hat{\psi}$ has the structure of a second-order V-statistic, we invoke asymptotic results from the theory of U- and V-statistics to establish the asymptotic normality of $\hat{\theta}_{\mathrm{Gini}}$.
For notational convenience, let $\psi^*$, $\mu^*$, and $\theta_{\mathrm{Gini}}^*$ denote the true values of
$\psi(F)$, $\mu(F)$, and $\theta_{\mathrm{Gini}}(F)$ defined in \eqref{Gini}, respectively.

\begin{theorem}
\label{Thm_gini} 
Suppose that Conditions C1--C5 in the Appendix are satisfied and that
\(
\int y^2 \rho^*(y)^{-1}\, dF(y) < \infty.
\)
Under model~\eqref{Alho}, as $N \to \infty$,
\[
N^{1/2}\big(\hat{\theta}_{\mathrm{Gini}} - \theta_{\mathrm{Gini}}^*\big)
\overset{d}{\to}
\mathcal{N}\big(0, \sigma_{\mathrm{Gini}}^2\big),
\]
where $\sigma_{\mathrm{Gini}}^2 = \bb_G^\top \bSigma_G \bb_G$, with
$\bb_G = (\mu^*)^{-1}\big(-\psi^*(\mu^*)^{-1},\, 1\big)^\top$, 
\[
\bSigma_G
=
\E\!\left\{\frac{\bxi_G(Y)\bxi_G(Y)^\top}{\rho^*(Y)}\right\}
+
\E\!\big\{\bxi_G(Y)\bv^\top(Y)\big\}
\bGamma
\E\!\big\{\bv(Y)\bxi_G^\top(Y)\big\},
\]
and
\[
\bxi_G(Y)
=
\left(
Y - \mu^*,\;
2\{YF(Y) + \int_Y^\infty t\, dF(t)\} - 2\psi^*
\right)^\top.
\]
\end{theorem}

\subsection{Semiparametric inference of inequality measures}
\label{Sec2.4}

In this section, we focus on constructing confidence intervals for the inequality measures under consideration.
Based on the asymptotic normality results established in Section~\ref{Sec2.3}, it is natural to construct Wald-type confidence intervals.
To estimate the asymptotic variances of the proposed estimators, we replace the unknown parameters with their consistent estimators.
Specifically, we estimate $(\balpha^*, \bbeta^*, \eta^*, F)$ by $(\hat{\balpha}, \hat{\bbeta}, \hat{\eta}, \hat{F})$, and estimate the density function $f(y)$ of $Y$ by the procedure as described below.

Since $Y$ is supported on $(0,\infty)$, we first estimate the density of the transformed variable $T=\log(Y)$, denoted by $k(t)$.
Given $\hat{F}(y)$ in \eqref{hatF}, $k(t)$ is estimated by
\[
\hat{k}(t)
=
\int K_b\big(t - \log(y)\big)\, d\hat{F}(y),
\]
where $K_b(x) = K(x/b)/b$ and $K(\cdot)$ is the standard normal kernel.
The bandwidth $b$ is selected according to
\[
b
=
1.06\, n^{-1/5}
\min\!\left(\widehat{\mathrm{IQR}}/1.34,\; \hat{\sigma}\right),
\]
where $\widehat{\mathrm{IQR}}$ and $\hat{\sigma}^2$ denote the interquartile range and variance estimators of $\log(Y)$, respectively, computed using $\hat{F}(y)$ \citep{Silverman1986}.
These quantities can be obtained using the methods described in Sections~\ref{Sec2.2} and~\ref{Sec2.3}.
Finally, the density of $Y$ is estimated by
\[
\hat{f}(y)
=
\hat{k}\big(\log(y)\big)/y.
\]

Let $\hat{\bOmega}(\tau_1,\tau_2)$ denote the estimator of $\bOmega(\tau_1,\tau_2)$ obtained by replacing
$(\balpha^*, \bbeta^*, \eta^*, F, f)$ with $(\hat{\balpha}, \hat{\bbeta}, \hat{\eta}, \hat{F}, \hat{f})$.
Similarly, let $(\hat{\sigma}_{\bu,h}^2, \hat{\sigma}_{\mathrm{Theil}}^2, \hat{\sigma}_{\mathrm{Gini}}^2)$ denote the estimators of
$(\sigma_{\bu,h}^2, \sigma_{\mathrm{Theil}}^2, \sigma_{\mathrm{Gini}}^2)$ obtained by replacing
$(\balpha^*, \bbeta^*, \eta^*, F)$ with $(\hat{\balpha}, \hat{\bbeta}, \hat{\eta}, \hat{F})$.
These variance estimators can be shown to be consistent.
By Slutsky's theorem, Wald-type confidence intervals can therefore be constructed for
$\theta_{\mathrm{quan}}^{\tau}(F)$,
$g(\theta_{\mathrm{quan}}^{\tau_1}, \theta_{\mathrm{quan}}^{\tau_2})$,
$\theta_{\bu,h}(F)$,
$\theta_{\mathrm{Theil}}(F)$, and
$\theta_{\mathrm{Gini}}(F)$.

For example, a $100(1-\alpha)\%$ Wald-type confidence interval for $\theta_{\mathrm{quan}}^{\tau}(F)$ is given by
\[
\left[
\hat{\theta}_{\mathrm{quan}}^{\tau}
-
\frac{z_{1-\alpha/2}\, \hat{\sigma}_{11}(\tau)}{\sqrt{N}},
\;
\hat{\theta}_{\mathrm{quan}}^{\tau}
+
\frac{z_{1-\alpha/2}\, \hat{\sigma}_{11}(\tau)}{\sqrt{N}}
\right],
\]
where $\hat{\sigma}_{11}(\tau)$ is the $(1,1)$ element of $\hat{\bOmega}(\tau,\tau)$, representing the estimated asymptotic standard deviation of $\hat{\theta}_{\mathrm{quan}}^{\tau}$, and $z_{1-\alpha/2}$ denotes the $100(1-\alpha/2)$th percentile of the standard normal distribution.
Confidence intervals for
$g(\theta_{\mathrm{quan}}^{\tau_1}, \theta_{\mathrm{quan}}^{\tau_2})$,
$\theta_{\bu,h}(F)$,
$\theta_{\mathrm{Theil}}(F)$, and
$\theta_{\mathrm{Gini}}(F)$
can be constructed analogously.
A simulation study evaluating the finite-sample performance of the proposed Wald-type confidence intervals for various inequality measures is presented in Section~\ref{Sec_simulation}.

\section{Expectation-Maximization algorithm} 
\label{sec.EM}

Since the $Y_i$'s are unobserved for nonrespondents, the problem can be viewed as a special case of missing data. The EM algorithm thus provides a natural framework for numerically computing $\hat{\bvarphi}$ defined in \eqref{hatphi}.

Recall that the observed data consist of
$\bO = \{(Y_i, D_i): i=1,\ldots,n\} \cup \{D_k = m+1: k=n+1,\ldots,N\}$.
Let
$\bO^* = \{(Z_k, D_k = m+1): k=n+1,\ldots,N\}$
denote the unobserved responses for the $N-n$ nonrespondents.
If both $\bO$ and $\bO^*$ were available, the complete-data likelihood would take the form
\bes
\prod_{i=1}^{n} \prod_{j=1}^m
\Big\{\rho_j(Y_i;\balpha,\bbeta)\, dF(Y_i)\Big\}^{I(D_i=j)}
\cdot
\prod_{k=n+1}^N
\{1-\rho(Z_k;\balpha,\bbeta)\}\, dF(Z_k).
\ees
Recall that $p_i = dF(Y_i)$ for $i=1,\ldots,n$, and that $F$ is modeled as
$F(y) = \sum_{i=1}^{n} p_i I(Y_i \le y)$.
Under this specification, each latent variable $Z_k$ can take values only in $\{Y_1,\ldots,Y_n\}$.
The corresponding complete-data log-likelihood is therefore
\ba
\label{ell_complete}
\ell_{\mathrm{EM}}(\bvarphi)
&=&
\sum_{i=1}^{n} \sum_{j=1}^m I(D_i=j)\log\{\rho_j(Y_i;\balpha,\bbeta)\}
+ \sum_{i=1}^{n}\log(p_i) \nonumber\\
&&
+ \sum_{i=1}^{n}\sum_{k=n+1}^N
I(Z_k = Y_i)
\big[\log(p_i) + \log\{1-\rho(Y_i;\balpha,\bbeta)\}\big],
\ea
subject to the constraint set $\mathcal{C}$ in \eqref{const}.

As in \citet{Dempster1977}, each iteration of the EM algorithm consists of two steps, the E-step and the M-step.
Suppose that $r$ iterations have been completed and the current parameter value is $\bvarphi^{(r)}$.
The $(r+1)$th iteration then proceeds as follows.

In the \emph{E-step}, given the current parameter value $\bvarphi^{(r)}$ and the observed data $\bO$, we compute the conditional expectation of the complete-data log-likelihood
\be
\label{Q_func}
Q(\bvarphi \mid \bvarphi^{(r)})
= \E\!\left\{\ell_{\mathrm{EM}}(\bvarphi) \mid \bO, \bvarphi^{(r)}\right\},
\ee
which is maximized in the subsequent M-step.
To evaluate $Q(\bvarphi \mid \bvarphi^{(r)})$, it suffices to compute
\be
\label{post}
\E\!\left\{ I(Z_k = Y_i) \mid \bO, \bvarphi^{(r)} \right\}
&= \Pr(Z_k = Y_i \mid Y_i, D_k = m+1, \bvarphi^{(r)})\\
&= \frac{p_i^{(r)}\{1-\rho(Y_i;\balpha^{(r)},\bbeta^{(r)})\}}{1-\eta^{(r)}},
\ee
where the final equality follows by an argument analogous to that leading to \eqref{part2.post}.
Combining \eqref{ell_complete}--\eqref{post}, we can write
\[
Q(\bvarphi \mid \bvarphi^{(r)})
=
\ell_1^{(r)}(\balpha,\bbeta)
+
\ell_2^{(r)}(p_1,\ldots,p_n),
\]
where
\bes
\ell_1^{(r)}(\balpha,\bbeta)
&=
\sum_{i=1}^{n}\sum_{j=1}^m I(D_i=j)\log\{\rho_j(Y_i;\balpha,\bbeta)\}
+ \sum_{i=1}^{n} w_i^{(r)} \log\{1-\rho(Y_i;\balpha,\bbeta)\}, \\
\ell_2^{(r)}(p_1,\ldots,p_n)
&=
\sum_{i=1}^{n} \big(w_i^{(r)} + 1\big)\log(p_i),
\ees
and
\[
w_i^{(r)}
=
(N-n)\,
\frac{p_i^{(r)}\{1-\rho(Y_i;\balpha^{(r)},\bbeta^{(r)})\}}
{1-\eta^{(r)}}.
\]

In the \emph{M-step}, we update the parameter vector from $\bvarphi^{(r)}$ to $\bvarphi^{(r+1)}$ by solving
\[
\bvarphi^{(r+1)}
=
\arg\max_{\bvarphi}
Q(\bvarphi \mid \bvarphi^{(r)})
\quad \text{subject to the constraints in \eqref{const}}.
\]
Owing to the additive structure of $Q(\bvarphi \mid \bvarphi^{(r)})$, the update $\bvarphi^{(r+1)}$ can be obtained explicitly as
\bes
p_i^{(r+1)}
&= \frac{w_i^{(r)} + 1}{N}, \quad i=1,\ldots,n, \nonumber\\
(\balpha^{(r+1)}, \bbeta^{(r+1)})
&=
\arg\max_{\balpha,\bbeta} \ell_1^{(r)}(\balpha,\bbeta), \\
\eta^{(r+1)}
&=
\sum_{i=1}^{n}
p_i^{(r+1)}
\rho\!\left(Y_i; \balpha^{(r+1)}, \bbeta^{(r+1)}\right).
\nonumber
\ees
It is worth noting that the objective function $\ell_1^{(r)}(\balpha,\bbeta)$ is proportional to the weighted log-likelihood of a logistic regression model.
Consequently, $(\balpha^{(r+1)}, \bbeta^{(r+1)})$ can be readily obtained by fitting a weighted logistic regression, which is supported by most standard statistical software.
Additional implementation details are provided in Section~4 of the Supplementary Material.

We iterate the E-step and M-step until the increase in the log-likelihood
$\ell(\bvarphi)$ in \eqref{log_lik} between successive iterations falls below a prespecified tolerance level, for example, $10^{-5}$.
The following proposition establishes the monotonicity property of the proposed EM algorithm.

\begin{proposition}
\label{nondecreasing}
For the EM algorithm described above, we have, for all $r \ge 0$,
\[
\ell\!\left(\bvarphi^{(r+1)}\right) \ge \ell\!\left(\bvarphi^{(r)}\right),
\]
that is, the log-likelihood function $\ell(\bvarphi)$ defined in \eqref{log_lik} is nondecreasing across successive EM iterations.

\end{proposition}

Based on extensive numerical experimentation, we find that the EM algorithm described above is not sensitive to the choice of initial values for $\bvarphi$.
Accordingly, in our numerical implementation, we initialize
$\balpha^{(0)} = \0$,
$\bbeta^{(0)} = \0$,
and $p_i^{(0)} = 1/n$ for $i = 1,\ldots,n$, and set
\[
\eta^{(0)}
=
\sum_{i=1}^n p_i^{(0)} \rho\!\left(Y_i; \balpha^{(0)}, \bbeta^{(0)}\right)
=
1 - 2^{-m}.
\]

\section{Simulation study}
\label{Sec_simulation}

In this section, we report simulation results assessing the finite-sample performance of the proposed semiparametric estimators and their associated confidence intervals for 
$\theta_{\text{quan}}^{\tau}(F)$ with $\tau = 0.25, 0.50,$ and $0.75$, 
$\theta_{\text{Theil}}(F)$ (as a representative member of the $\theta_{\mathbf{u},h}(F)$ class), 
and $\theta_{\text{Gini}}(F)$ under nonignorable nonresponse.

Three distributions are considered for $F(y)$: $\mathrm{Exp}(1)$, $\chi^2(1.5)$, and $\mathrm{Gam}(0.8,0.25)$, where $\mathrm{Exp}(a_1)$ denotes an exponential distribution with rate parameter $a_1$, $\chi^2(a_2)$ denotes a chi-squared distribution with $a_2$ degrees of freedom, and $\mathrm{Gam}(a_3,a_4)$ denotes a gamma distribution with shape parameter $a_3$ and rate parameter $a_4$. The callback indicator $D$ is generated according to Alho's callback model~\eqref{Alho} with $\bq(y)=\log(y)$ and $m=2$. The parameter vector $(\balpha^\top,\beta)$ is set to $(-1.5,\,0.5,\,-0.5)$, yielding moderate response rates across contacts. The negative value of $\beta$ implies that the response probability decreases as $Y$ increases.

The resulting true response probabilities for each contact attempt and the true values of the inequality measures are presented in Table~\ref{true_value}. Simulations are conducted for sample sizes $N = 1000$ and $2000$, with $M = 5000$ Monte Carlo replications for each design.

\begin{table}[!h]
\tabcolsep 8pt
\small
\renewcommand\arraystretch{1}
\caption{True response probabilities at each contact, overall nonresponse rate, and inequality measures under the simulation design.
}
\label{true_value}
\centering
\begin{tabular}{l ccc ccccc}
\toprule
$F(y)$ & $\Pr(D=1)$ & $\Pr(D=2)$ & $1-\eta^*$ & $\theta_\text{quan}^{0.25*}$ & $\theta_\text{quan}^{0.50*}$ & $\theta_\text{quan}^{0.75*}$ & $\theta_{\text{Theil}}^*$ & $\theta_{\text{Gini}}^*$ \\ 
\midrule
$\text{Exp}(1)$         & 0.245 & 0.494 & 0.261 & 0.288 & 0.693 & 1.386 &  0.423 & 0.500 \\
$\chi^2(1.5)$           & 0.236 & 0.473 & 0.291 & 0.307 & 0.908 & 2.068 &  0.535 & 0.556 \\
$\text{Gam}(0.8,0.25)$  & 0.175 & 0.443 & 0.381 & 0.713 & 2.005 & 4.424 & 0.508 & 0.544 \\
\bottomrule
\end{tabular}
\end{table}

\subsection{Performance of point estimation}

We compare the finite-sample performance of the following three estimators for the inequality measures:
\begin{itemize}
\setlength{\itemsep}{0pt}
\setlength{\parsep}{0pt}
\setlength{\parskip}{0pt}
\item \text{CC}: a naive estimator based on complete cases only;
\item \text{Proposed}: the proposed semiparametric full-likelihood estimator;
\item \text{Ideal}: a benchmark estimator computed using the full data.
\end{itemize}
The CC estimator discards all units with nonresponse or missing values and is therefore generally inconsistent under nonignorable nonresponse. The Ideal estimator is included solely as a benchmark and is not available in practice.

Finite-sample performance of a point estimator is evaluated using the relative bias (RB) and the root mean squared error (RMSE), defined as
\begin{equation*}
\text{RB} = \frac{1}{M} \sum_{i=1}^{M} \frac{\hat{x}^{(i)} - x}{x},
\qquad
\text{RMSE} = \left\{ \frac{1}{M} \sum_{i=1}^{M} \big(\hat{x}^{(i)} - x\big)^2 \right\}^{1/2},
\end{equation*}
where $\hat{x}^{(i)}$ denotes the estimate of a target parameter with true value $x$ in the $i$th replication $(i=1,\ldots,M)$. Simulation results are reported in Table~\ref{simu_1}.

\begin{table}[!ht]
\tabcolsep 1.6pt
\small
\renewcommand\arraystretch{1}
\caption{Relative bias (RB) $\times 100$ and root mean squared error (RMSE) $\times 100$ for quantiles, the Theil index, and the Gini index under the simulation design.}
\centering
\label{simu_1}
\begin{tabular}{l l cccccc cccc}
\toprule
 &  & \multicolumn{2}{c}{$\theta_\text{quan}^{0.25}(F)$} & \multicolumn{2}{c}{$\theta_\text{quan}^{0.50}(F)$} & \multicolumn{2}{c}{$\theta_\text{quan}^{0.75}(F)$} & \multicolumn{2}{c}{$\theta_{\text{Theil}}(F)$} & \multicolumn{2}{c}{$\theta_{\text{Gini}}(F)$} \\
$F(y)$ & method & RB & RMSE & RB & RMSE & RB & RMSE & RB & RMSE & RB & RMSE \\
\midrule
 & & \multicolumn{10}{c}{$N=1000$}\\
\cline{3-12}
Exp$(1)$        & CC    & -20.520 & 6.156 & -17.736 & 12.689 &-14.959 & 21.486 &  7.231 & 3.731 & 3.424 & 2.028\\
             & Proposed &   0.248 & 2.050 &   0.193 &  3.908 &  0.146 &  7.441 & -0.152 & 2.187 & 0.188 & 1.146\\
                & Ideal &   0.361 & 1.831 &   0.097 &  3.144 & -0.022 &  5.456 & -0.079 & 1.700 & 0.114 & 0.913\\ 
$\chi^2(1.5)$   & CC    & -30.928 & 9.715 & -26.520 & 24.543 &-22.098 & 46.660 & 10.291 & 6.208 & 4.548 & 2.794\\
             & Proposed &   0.238 & 2.695 &   0.187 &  6.283 &  0.037 & 13.213 & -0.304 & 2.971 & 0.131 & 1.305\\
                & Ideal &   0.348 & 2.430 &   0.056 &  4.970 & -0.147 &  9.510 & -0.142 & 2.127 & 0.069 & 0.963\\
Gam$(0.8,0.25)$& CC     & -36.973 & 26.789 & -31.352 & 63.668 & -25.822 & 115.994 & 13.871 & 7.654 & 6.112 & 3.552\\
             & Proposed &   0.237 &  6.577 &   0.234 & 15.337 &   0.127 &  32.241 & -0.490 & 3.085 & 0.129 & 1.401\\
                & Ideal &   0.368 &  5.383 &   0.041 & 10.493 &  -0.149 &  19.644 & -0.160 & 2.037 & 0.064 & 0.958\\
\midrule
 & & \multicolumn{10}{c}{$N=2000$}\\
\cline{3-12}
Exp$(1)$        & CC    & -20.807 & 6.112 & -17.894 & 12.606 & -15.052 & 21.250 &  7.396 & 3.472 & 3.392 & 1.860\\
            & Proposed  &   0.031 & 1.445 &  -0.037 &  2.809 &  -0.024 &  5.205 & -0.002 & 1.559 & 0.124 & 0.815\\
                & Ideal &   0.129 & 1.285 &   0.029 &  2.260 &  -0.044 &  3.894 & -0.038 & 1.214 & 0.057 & 0.653\\
$\chi^2(1.5)$   & CC    & -31.159 & 9.675 & -26.634 & 24.413 & -22.054 & 46.079 & 10.291 & 6.208 & 4.504 & 2.637\\
            & Proposed  &   0.118 & 1.932 &   0.038 &  4.417 &  -0.052 &  9.168 & -0.304 & 2.971 & 0.058 & 0.900\\
                & Ideal &   0.179 & 1.721 &  -0.004 &  3.499 &  -0.111 &  6.684 & -0.142 & 2.127 & 0.033 & 0.675\\
Gam$(0.8,0.25)$& CC     & -37.140 & 26.686 & -31.459 & 63.490 & -25.781 & 114.937 & 14.119 & 7.469 & 6.092 & 3.425\\
             & Proposed &   0.102 &  4.659 &  -0.011 & 10.611 &  -0.019 &  22.355 & -0.161 & 2.168 & 0.097 & 0.979\\
                & Ideal &   0.203 &  3.796 &  -0.023 &  7.395 &  -0.112 &  13.714 & -0.084 & 1.423 & 0.029 & 0.669\\
\bottomrule
\end{tabular}
\end{table}

We summarize the main findings from Table~\ref{simu_1}. First, as anticipated, the CC estimator performs poorly under nonignorable nonresponse, exhibiting substantial RB and RMSE across all simulation settings. Moreover, the CC estimator does not display consistency, as neither RB nor RMSE decreases when the sample size increases from $N=1000$ to $N=2000$. The CC estimator systematically underestimates the quantiles (negative RB) and overestimates the Theil and Gini indices (positive RB).
Second, both the proposed semiparametric estimator and the Ideal estimator exhibit negligible RB across all settings. The RMSEs of the proposed estimator are comparable to those of the benchmark Ideal estimator and decrease as the sample size increases.
Overall, the results indicate that the proposed semiparametric approach effectively corrects for bias induced by nonignorable nonresponse and achieves near-benchmark efficiency in finite samples.

\subsection{Performance of confidence intervals}

The finite-sample performance of confidence intervals is evaluated using the coverage probability (CP), reported as a percentage, and the average length (AL), defined as
\begin{equation*}
\text{CP} = \frac{1}{M} \sum_{i=1}^{M} I\{x \in \mathcal{I}^{(i)}\},
\qquad
\text{AL} = \frac{1}{M} \sum_{i=1}^{M} \left| \mathcal{I}^{(i)} \right|,
\end{equation*}
where $\mathcal{I}^{(i)}$ denotes the confidence interval for a target parameter $x$ in the $i$th replication, and $|\mathcal{I}^{(i)}|$ denotes its length. Simulation results for the nominal 95\% confidence intervals of the quantiles, the Theil index, and the Gini index described in Section~\ref{Sec2.4}, for sample sizes $N=1000$ and $2000$, are reported in Table~\ref{simu_ci}.

\begin{table}[!ht]
\tabcolsep 6.8pt
\small
\renewcommand\arraystretch{1}
\caption{Coverage probability (CP) and average length (AL) of 95\% confidence intervals for quantiles, the Theil index, and the Gini index under the simulation design.
}
\centering
\label{simu_ci}
\begin{tabular}{l cccccc cccc}
\toprule
 & \multicolumn{2}{c}{$\theta_\text{quan}^{0.25}(F)$} & \multicolumn{2}{c}{$\theta_\text{quan}^{0.50}(F)$} & \multicolumn{2}{c}{$\theta_\text{quan}^{0.75}(F)$} & \multicolumn{2}{c}{$\theta_{\text{Theil}}(F)$} & \multicolumn{2}{c}{$\theta_{\text{Gini}}(F)$} \\
$F(y)$ & CP & AL & CP & AL & CP & AL & CP & AL & CP & AL \\
\midrule
 & \multicolumn{10}{c}{$N=1000$}\\
 \cline{2-11}
Exp$(1)$      & 0.948 & 0.080 & 0.958 & 0.160 & 0.958 & 0.308 & 0.938 & 0.084 & 0.947 & 0.045\\
$\chi^2(1.5)$ & 0.947 & 0.106 & 0.955 & 0.255 & 0.966 & 0.561 & 0.921 & 0.110 & 0.943 & 0.049\\
Gam$(0.8,0.25)$& 0.951 & 0.258 & 0.960 & 0.624 & 0.964 & 1.343 & 0.923 & 0.115 & 0.942 & 0.054\\
\midrule
 & \multicolumn{10}{c}{$N=2000$}\\
 \cline{2-11}
Exp$(1)$       & 0.948 & 0.056 & 0.954 & 0.112 & 0.960 & 0.215 & 0.943 & 0.060 & 0.946 & 0.032\\
$\chi^2(1.5)$  & 0.950 & 0.075 & 0.959 & 0.179 & 0.967 & 0.390 & 0.939 & 0.080 & 0.948 & 0.035\\
Gam$(0.8,0.25)$& 0.949 & 0.183 & 0.960 & 0.437 & 0.959 & 0.933 & 0.941 & 0.084 & 0.949 & 0.039\\
\bottomrule
\end{tabular}
\end{table}

We summarize the main findings from Table~\ref{simu_ci}. Overall, the confidence intervals constructed using the proposed semiparametric method achieve coverage probabilities close to the nominal 95\% level across most settings, with moderate average lengths. Some undercoverage is observed for certain parameters, such as $\theta_{\text{Theil}}(F)$, when the sample size is $N=1000$. As the sample size increases to $N=2000$, coverage probabilities approach the nominal level and interval lengths become shorter. These results indicate that the proposed semiparametric inference procedure provides reliable finite-sample inference under nonignorable nonresponse with callback data.

\section{A real data example}
\label{Sec_real}

In this section, we apply the proposed estimation and inference procedure to data from the Consumer Expenditure Survey (CES), which is conducted by the United States Bureau of Labor Statistics to collect detailed information on household expenditures and related characteristics. A comprehensive description of the survey design is available at \url{https://www.census.gov/programs-surveys/ce.html}.

Our analysis uses the public-use adult data and associated paradata from the first quarter of the 2024 CES. For illustration, we consider the total amount of family resources after taxes over the previous 12 months (measured in \$10,000) as the study variable $Y$. The sample consists of 10,273 households with positive recorded values, with an overall nonresponse rate of 54.54\%. The objective is to conduct efficient and reliable statistical inference for distributional and inequality measures, including quantiles, the Theil index, and the Gini index, using the CES data.

\begin{figure}[!ht]
    \centering
    \setlength{\abovecaptionskip}{0pt}
    \setlength{\belowcaptionskip}{0pt}
    \includegraphics[width=0.8\linewidth]{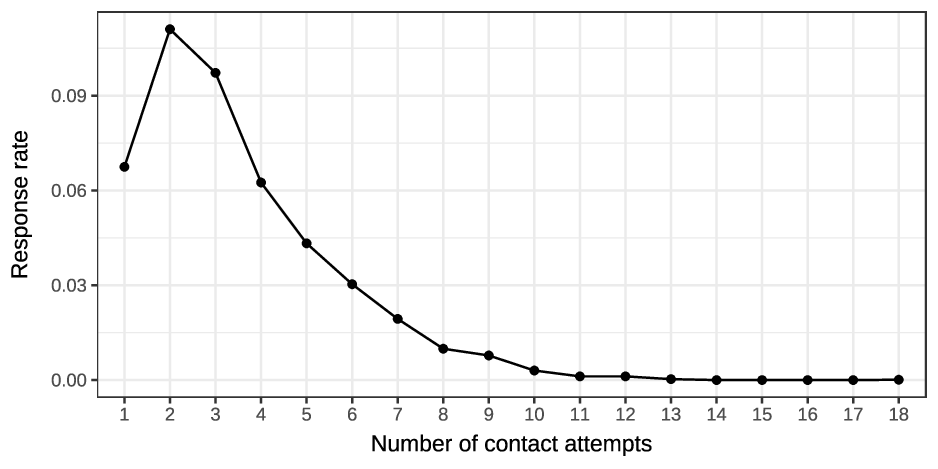}
    \caption{Response rate by number of contact attempts in the CES data.}
    \label{fig.rate}
\end{figure}

The maximum number of contact attempts recorded in the CES paradata is 18. Figure~\ref{fig.rate} displays the response rates by number of contact attempts. The callback strategy is most effective during the initial contacts, with response rates declining as the number of calls increases. Specifically, 6.75\% and 11.10\% of households responded on the first and second calls, respectively, while an additional 27.61\% responded in subsequent calls, yielding an overall response rate of 45.46\%.
Motivated by this pattern, we define four callback categories: $D=1$ and $D=2$ for households responding on the first and second calls, respectively; $D=3$ for those responding at later calls; and $D=4$ for households that did not respond.

To implement the proposed method, we use the Akaike Information Criterion (AIC) and the Bayesian Information Criterion (BIC) to select the functional form of $\bq(y)$ in model~\eqref{Alho} from the candidate set $\{y, y^2, \log(y), \log^2(y)\}$ and their combinations. Based on the results in Table~\ref{AIC_BIC}, $\bq(y)=\log(y)$ is selected, as it yields the smallest AIC and BIC values. The semiparametric full-likelihood estimates of the model parameters and their corresponding 95\% confidence intervals are reported in Table~\ref{tab.model.est.realdata}. In particular, the estimate of $\beta$ is $-0.191$, with a 95\% confidence interval of $[-0.293,-0.090]$, indicating statistically significant evidence of nonignorable nonresponse in the CES data, with response propensity decreasing as the value of $Y$ increases.

\begin{table}[!htt]
\small
\renewcommand\arraystretch{1}
\tabcolsep 45pt
\caption{AIC and BIC values for different choices of $\bq(y)$ in the callback model \eqref{Alho} based on CES data.}
\label{AIC_BIC}
\centering
\begin{tabular}{lcc}
\toprule
$\bq(y)$ & AIC & BIC \\
\midrule
    $y$ & 101900.42 & 101936.61 \\
    $y^2$ & 101902.26 & 101938.45 \\
    $\log(y)$ & \emph{101899.09} & \emph{101935.29}\\
    $\log(y)^2$ & 101902.29 & 101938.49\\
    $(y, y^2)^\top$ & 101902.31 & 101945.74 \\
    $(y, \log(y))^\top$ & 101900.18 & 101943.61\\
    $(y, \log(y)^2)^\top$ & 101901.70 & 101945.13\\
    $(y^2, \log(y))^\top$ & 101899.80 & 101943.20\\
    $(y^2, \log(y)^2)^\top$ & 101903.85 & 101947.28\\
    $(\log(y), \log(y)^2)^\top$ & 101900.60 & 101944.00\\
    $(y, y^2, \log(y))^\top$ & 101901.67 & 101952.34\\
    $(y, y^2, \log(y)^2)^\top$ & 101903.52 & 101954.19\\
    $(y, \log(y), \log(y)^2)^\top$ & 101902.18 & 101952.85\\
    $(y, y^2, \log(y), \log(y)^2)^\top$ & 101903.00 & 101960.90\\
\bottomrule
\end{tabular}
\end{table}

\begin{table}[!ht]
\small
\renewcommand\arraystretch{1}
\tabcolsep 4.5pt
\caption{Semiparametric estimates of $(\balpha,\beta,\eta)$ and corresponding 95\% confidence intervals based on CES data.}
\label{tab.model.est.realdata}
\centering
\begin{tabular}{lccccc}
\toprule
& $\alpha_1$ & $\alpha_2$ & $\alpha_3$ & $\beta$ &  $\eta$ \\
\midrule
Estimate & $-2.278$ & $-1.647$ & $-0.313$  & $-0.191$ & $0.455$\\
95\% CI & $[-2.480, -2.076]$ & $[-1.849, -1.445]$ & $[-0.524, -0.103]$ & $[-0.293, -0.090]$ & [0.445, 0.464]\\
\bottomrule
\end{tabular}
\end{table}

The proposed semiparametric full-likelihood estimates of the quantiles, the Theil index, and the Gini index, together with their 95\% confidence intervals, are reported in Table~\ref{tab.inequality.est.realdata}. For comparison, the table also includes the corresponding estimates based on the 4{,}670 complete cases. The CC estimates differ markedly from the proposed semiparametric estimates, and all CC estimates fall outside the 95\% confidence intervals constructed using the proposed method. This finding is consistent with the simulation results and indicates that the observed complete cases may not represent a random sample of the target distribution.

\begin{table}[!ht]
\small
\renewcommand\arraystretch{1}
\tabcolsep 29pt
\caption{Estimates of quantiles, the Theil index, and the Gini index with 95\% confidence intervals based on CES data.}
\label{tab.inequality.est.realdata}
\centering
\begin{tabular}{l ccc}
\toprule
Inequality measure & CC & Proposed  & 95\% CI \\
\midrule
$\theta_\text{quan}^{0.25}(F)$ & 3.470  &  3.930 & $[ 3.641, 4.218]$ \\
$\theta_\text{quan}^{0.50}(F)$ & 6.737  &  7.435 & $[ 6.987, 7.882]$ \\
$\theta_\text{quan}^{0.75}(F)$ & 11.719 & 12.873 & $[12.156,13.590]$ \\
$\theta_{\text{Theil}}(F)$ &  0.349 & 0.332 & [0.317, 0.347] \\	
$\theta_{\text{Gini}}(F)$  &  0.455 & 0.445 & [0.436, 0.454] \\
\bottomrule
\end{tabular}
\end{table}

\begin{figure}[h!]
    \centering
    \setlength{\abovecaptionskip}{0pt}
    \setlength{\belowcaptionskip}{0pt}
    \includegraphics[width=0.8\linewidth]{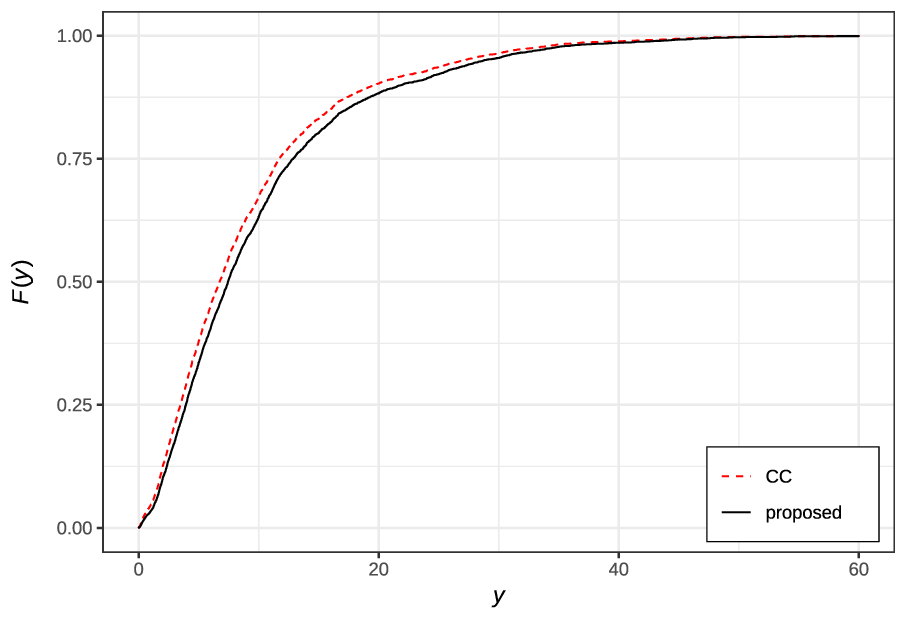}
    \caption{Estimated distribution function $F(y)$ based on CES data using the proposed semiparametric method (solid line) and the complete-case method (dotted line).}
    \label{fig.Fhat}
\end{figure}

Figure~\ref{fig.Fhat} compares the estimated distribution functions obtained from the proposed method and the CC method. A pronounced discrepancy between the two estimates is evident. Overall, these results suggest that ignoring nonignorable nonresponse can lead to substantial bias and unreliable inference for inequality measures, whereas the proposed semiparametric approach, which effectively incorporates callback information, yields more reliable estimates and valid inference for a range of commonly used inequality measures.

\section{Concluding remarks}
\label{Sec_discussion}

Nonignorable nonresponse has long been recognized as a major challenge in survey data analysis, yet its impact on inference for inequality measures remains relatively underexplored. In this paper, we develop a unified semiparametric framework for estimation and inference of inequality measures in the presence of nonignorable nonresponse. The proposed approach leverages callback data within a semiparametric full-likelihood formulation, leaving the underlying distribution $F(y)$ unspecified.
We establish the asymptotic properties of the proposed estimators for a broad class of inequality measures, enabling valid inference under nonignorable nonresponse. From a practical perspective, we introduce an efficient and numerically stable EM algorithm with theoretical guarantees, facilitating straightforward implementation. Simulation studies and an empirical application illustrate the favorable finite-sample performance of the proposed estimation and inference procedures for commonly used inequality measures.

Several directions merit further investigation. First, in addition to nonignorable nonresponse, survey data often exhibit other forms of incompleteness, such as grouped reporting \citep{Gastwirth1972} and top-coding \citep{Feng2006}. Extending the proposed framework to accommodate these features would be of practical interest. Second, it would be worthwhile to study inference for more general parameters defined through moment restrictions \citep{Qin1994} under nonignorable nonresponse. Finally, alternative sources of auxiliary information, such as regional response rates \citep{Korinek2006,Korinek2007}, have been proposed as substitutes for callback data. Developing semiparametric inference theory under such settings represents another promising direction for future research.

\section*{Declaration of competing interest}
The authors declare that they have no known competing financial interests or personal relationships that could have appeared to influence the work reported in this paper.

\section*{Acknowledgments}
C.~Wang's work was supported by the Humanities and Social Sciences Foundation of the Ministry of Education of China (24YJA910005) and the National Natural Science Foundation of China (71988101); T.~Yu's work is supported in part by the Singapore Ministry of Education Academic Research Tier 1 Fund (A-8000413-00-00); P.~Li's work was supported by the Natural Sciences and Engineering Research Council of Canada (RGPIN-2020-0496).

\appendix
\section{Form of $\bGamma$ and regularity conditions}
Recall that the estimator $\hat{\bphi}$ of the unknown parameter vector is obtained by maximizing the log-likelihood function in \eqref{log_lik} subject to the constraint set $\mathcal{C}$ in \eqref{const}. This maximization can be carried out in two steps. We first profile out the $p_i$'s using the Lagrange multiplier method \citep{Qin1994}. Specifically, this yields
\begin{equation}
\label{lagrange.pi}
p_i = \frac{1}{n}\,\frac{1}{1 + \lambda\{\rho(Y_i; \balpha, \bbeta) - \eta\}},
\qquad i = 1,\ldots,n,
\end{equation}
where the Lagrange multiplier $\lambda=\lambda(\balpha,\bbeta,\eta)$ solves
\begin{equation}
\label{score_lambda}
\sum_{i=1}^{n}
\frac{\rho(Y_i; \balpha, \bbeta) - \eta}
     {1 + \lambda\{\rho(Y_i; \balpha, \bbeta) - \eta\}}
= 0.
\end{equation}
Substituting \eqref{lagrange.pi} into the log-likelihood function in \eqref{log_lik} and ignoring additive constants, the resulting profile semiparametric full log-likelihood is
\[
\label{prof_log_lik}
\ell_{\mathrm{p}}(\balpha, \bbeta, \eta)
= \sum_{i=1}^{n} \sum_{j=1}^{m} I(D_i = j)\log\{\rho_j(Y_i; \balpha, \bbeta)\}
- \sum_{i=1}^{n} \log\!\Big[1 + \lambda\{\rho(Y_i; \balpha, \bbeta) - \eta\}\Big]
+ (N - n)\log(1 - \eta).
\]
The semiparametric estimators $(\hat{\balpha}, \hat{\bbeta}, \hat{\eta})$ are defined as the maximizer of $\ell_{\mathrm{p}}(\balpha, \bbeta, \eta)$.

Let $\bnu = (\balpha^\top, \bbeta^\top, \eta, \lambda)^\top$ and define 
\[
H(\bnu)
= \sum_{i=1}^{n} \sum_{j=1}^{m} I(D_i = j)\log\{\rho_j(Y_i; \balpha, \bbeta)\}
- \sum_{i=1}^{n} \log\!\Big[1 + \lambda\{\rho(Y_i; \balpha, \bbeta) - \eta\}\Big]
+ (N - n)\log(1 - \eta).
\]
Define the individual contribution
\begin{equation*}
H_i(\bnu)
=
\sum_{j=1}^{m} I(D_i = j)\log\{\rho_j(Y_i; \balpha, \bbeta)\}
- I(D_i \le m)\log\!\Big[1 + \lambda\{\rho(Y_i; \balpha, \bbeta) - \eta\}\Big]
+ I(D_i = m+1)\log(1 - \eta).
\end{equation*}
Then $H(\bnu)$ can be written as
\(
H(\bnu) = \sum_{i=1}^{N} H_i(\bnu).
\)
Let $\hat{\lambda}$ denote the solution to \eqref{score_lambda} evaluated at $(\balpha, \bbeta, \eta) = (\hat{\balpha}, \hat{\bbeta}, \hat{\eta})$, and define
\(
\hat{\bnu} = (\hat{\balpha}^\top, \hat{\bbeta}^\top, \hat{\eta}, \hat{\lambda})^\top.
\)
It follows that
\[
\frac{\partial H(\hat{\bnu})}{\partial \bnu} = \mathbf{0}.
\]

Let
\(
\bnu^* = \big(\balpha^{*\top},\, \bbeta^{*\top},\, \eta^*,\, 1/\eta^*\big)^\top.
\)
To derive the asymptotic distribution of $\hat{\bnu}$, we first obtain a linear approximation to $\hat{\bnu} - \bnu^*$. This requires defining
\[
\bV = -N^{-1}\, \E\!\left\{\frac{\partial^2 H(\bnu^*)}{\partial \bnu\, \partial \bnu^\top}\right\}.
\]
Based on the structure of $\bV$, the matrix $\bGamma$ appearing in Theorem~\ref{Thm_F} is defined as
\begin{equation}
\label{Gamma}
\bGamma = \bV^{-1} + \bV^{-1}\bM\,\bV^{-1},
\end{equation}
where
\[
\bM =
\begin{pmatrix}
    \0_{m+d}\0_{m+d}^\top & \0_{m+d} & \0_{m+d} \\
    \0_{m+d}^\top        & 2(1-\eta^*)^{-1} & -\eta^* \\
    \0_{m+d}^\top        & -\eta^*          & 0
\end{pmatrix}.
\]

The asymptotic results in Section~\ref{Sec2.3} are established under the following regularity conditions.

\begin{itemize}
\item[C1.] The components of $\bq(y)$ in model~\eqref{Alho} are stochastically linearly independent.

\item[C2.] The true response probability satisfies $\eta^* = \Pr(D \le m) \in (0,1)$.

\item[C3.] For each $j = 1,\ldots,m$,
\[
\int_0^\infty \exp\{\alpha_j + \bbeta^\top \bq(y)\}\, dF(y) < \infty
\]
for all $(\balpha,\bbeta)$ in a neighborhood of the true parameter values $(\balpha^*, \bbeta^*)$.

\item[C4.] All elements of $\bV$ exist and are finite. Moreover, $\bV$ is nonsingular.

\item[C5.] There exist a positive constant $\epsilon$ and a function $L(Y,D)$ with finite expectation such that, for any three components $\nu_{l_1}$, $\nu_{l_2}$, and $\nu_{l_3}$ of $\bnu$,
\[
\sup_{\|\bnu - \bnu^*\| \le \epsilon}
\left|
\frac{\partial^3 H_i(\bnu)}
     {\partial \nu_{l_1}\,\partial \nu_{l_2}\,\partial \nu_{l_3}}
\right|
\le L(Y_i, D_i),
\]
where $\|\cdot\|$ denotes the Euclidean norm.
\end{itemize}

Condition C1 ensures identifiability of the model parameters. Condition C2 excludes degenerate cases in which the survey exhibits either no response or full response. Under Condition C3, the moment generating function of $\bq(y)$ with respect to the true distribution $F(y)$ exists in a neighborhood of the true parameter values $(\balpha^*, \bbeta^*)$, implying that all finite-order moments of $\bq(y)$ are finite. Conditions C3--C5 together ensure the validity of a quadratic approximation to $H(\bnu)$.

\bibliographystyle{elsarticle-harv}
\bibliography{callback}

\end{document}